%% file: sn-article.tex
\documentclass[sn-mathphys,Numbered]{sn-jnl}

\usepackage{graphicx}%
\usepackage{multirow}%
\usepackage{amsmath,amssymb,amsfonts}%
\usepackage{amsthm}%
\usepackage{mathrsfs}%
\usepackage[title]{appendix}%
\usepackage{xcolor}%
\usepackage{textcomp}%
\usepackage{manyfoot}%
\usepackage{booktabs}%
\usepackage{algorithm}%
\usepackage{algorithmicx}%
\usepackage{algpseudocode}%
\usepackage{listings}%
\usepackage{cleveref}
\usepackage{mathtools}
\usepackage{bbm}
\usepackage{pgfplots}
\usepgfplotslibrary{groupplots}
\usetikzlibrary{matrix}
\pgfplotsset{compat=1.17} 

\newtheorem{definition}{Definition}%
\newtheorem{problem}{Problem}

\def\R{\mathbb{R}}

\DeclareMathOperator*{\QRE}{QRE}
\DeclareMathOperator*{\softmax}{softmax}

\begin{document}

\title[relationship game]{Relationship Design for Socially-Aware Behavior in Static Games}


\author[1]{\fnm{Shenghui} \sur{Chen}}\email{shenghui.chen@utexas.edu}

\author[1]{\fnm{Yigit} \spfx{E.} \sur{Bayiz}}\email{egebayiz@utexas.com}

\author[1]{\fnm{David} \sur{Fridovich-Keil}}\email{dfk@utexas.com}

\author[1]{\fnm{Ufuk} \sur{Topcu}}\email{utopcu@utexas.com}

\affil*[1]{\orgname{University of Texas at Austin}}


\abstract{
Autonomous agents can adopt socially-aware behaviors to reduce social costs, mimicking the way animals interact in nature and humans in society.
We present a new approach to model socially-aware decision-making that includes two key elements: bounded rationality and inter-agent relationships. We capture the inter-agent relationships by introducing a novel model called a relationship game and encode agents' bounded rationality using quantal response equilibria.
For each relationship game, we define a social cost function and formulate a mechanism design problem to optimize weights for relationships that minimize social cost at the equilibrium.
We address the multiplicity of equilibria by presenting the problem in two forms: Min-Max and Min-Min, aimed respectively at minimization of the highest and lowest social costs in the equilibria.
We compute the quantal response equilibrium by solving a least-squares problem defined with its Karush-Kuhn-Tucker conditions, and propose two projected gradient descent algorithms to solve the mechanism design problems. Numerical results, including two-lane congestion and congestion with an ambulance, confirm that these algorithms consistently reach the equilibrium with the intended social costs.
}


\maketitle

\input{sections/1_introduction}
\input{sections/2_related}
\input{sections/3_preliminaries}
\input{sections/4_problem}
\input{sections/5_method}
\input{sections/6_numerical}
\input{sections/7_conclusion}
\section{Acknowledgement}
The authors would like to thank Yue Yu for his insights.

\bibliography{ref}

\begin{appendices}
\input{sections/appendix_inverse}
\end{appendices}

\end{document}

%% file: sections/1_introduction.tex
\section{Introduction}\label{sec:intro}
From the smallest insects to the largest mammals, natural agents demonstrate a remarkable capacity for socially-aware decision-making, resulting in behaviors ranging from competition to altruism. For example, bees exhibit highly cooperative and altruistic behavior in colonies, from foraging for resources to prioritizing the bee queen, because they act in the interest of the hive's collective needs instead of individual gains \cite{naeger2013altruistic}.

As autonomous agents become more pervasive, we expect them to be capable of co-existing with humans, creating a need for them to exhibit a similar kind of social awareness. Socially-aware decision-making has the added benefit of potentially improving the efficiency of the system.
Consider a flow of traffic with autonomous vehicles navigating alongside human-operated ones, which may make unpredictable or less-than-optimal choices. Purely rational autonomous vehicles, lacking in social awareness, would struggle to reduce traffic congestion, as this task requires them to take into account the complex web of social relationships between each other and human-operated vehicles.

General-sum static games offer a mathematical formalism to capture the interaction among multiple agents. 
In these games, agents are typically modeled as choosing Nash equilibrium strategies where no agent can reduce their cost by changing their strategy unilaterally \cite{bacsar1998dynamic}.
However, real-world scenarios often involve bounded rationality, where agents do not always choose optimal strategies. 
In such scenarios, instead of Nash equilibria, we model agent responses with \textit{quantal response equilibria}, where strategies are probabilistically chosen based on potential costs~\cite{mckelvey1995quantal}.

Besides modeling the bounded rationality of agents, socially-aware decision-making also involves the consideration of social relationships.
Humans are able to trust each other and behave cooperatively without much training, achieving higher efficiency because they are guided by ethical principles \cite{kuipers2020perspectives}. One important ethical principle is altruism---acting in a way that cares about and benefits others, but how do people calibrate how much they should care about other individuals in society? Kleiman-Weiner et al. \cite{kleiman2017learning} propose that the amount of care over specific people should be determined by abstract relationships. For example, people usually care more about others in their family than strangers. We hypothesize that autonomous agents can similarly attain socially-aware behaviors by introducing relationships among them.
These social relationships, integral to this framework, can be effectively modeled as graphs. By representing agents as nodes and their relationships as edges, we can map out the possible social interactions among agents.
This graph-based approach enables us to model how much each agent should take the costs of other agents into account.

In this paper, we present a novel approach to model socially-aware decision-making within a static game framework. 
Social scenarios often have inherent symmetries that dictate the range of relationships that can exist between agents. 
For example in a traffic setting, a regular car has a different relationship to an ambulance compared to another regular car. 
We associate each of these relationships with an adjacency matrix. Assigning weights to each relationship transforms the game's structure, resulting in a game where an agent's cost is influenced by the costs of others they are related to.

Our aim in this paper is to find the optimal weight assignment that obtains minimal social cost. Due to the multiplicity of quantal response equilibria, we formulate this goal via two bi-level optimization problems: Min-Max and Min-Min. We then propose two projected gradient descent algorithms to solve these two problems and empirically validate the algorithms on two different congestion game scenarios.


%% file: sections/2_related.tex
\section{Related Work}\label{sec:related}
{\noindent \bf Altruism in Games: } Cultivating altruistic behavior in static games is often challenging as agents face a dilemma where minimizing social costs does not align with optimizing their individual cost function. This dichotomy has led to the definitions of \textit{price of anarchy} and \textit{price of stability} \cite{roughgarden2010algorithmic}. 
The price of anarchy refers to the social cost ratio between taking optimal actions and the Nash equilibrium with maximal social cost. Conversely, the price of stability is the social cost ratio between optimal actions and the Nash equilibrium with minimal social cost. 
These metrics are by far the most commonly used ones for analyzing the cost of selfish behavior in the overall game performance. Several papers have proposed tight bounds on the price of anarchy in the context of atomic congestion games and cost-sharing games \cite{chen2014poa, caragiannis2010altruismcongestion, gollapudi2017profitsharing}. There are also works that investigate the price of anarchy \cite{bhwalkar2014weightedcongestion, gairing2020weightedpos} and the price of stability \cite{gairing2020weightedpos} in weighted congestion games. In this paper, we focus on the problem of finding the optimal cost-sharing mechanism in static games, and propose parameter optimization methods to minimize the price of anarchy and the price of stability by encouraging cost-sharing between agents. We show the performance of these methods on weighted congestion games.

{\noindent \bf Differentiable Optimization: } In this paper we optimize the game parameters based on the game solution they induce. Unfortunately, both Nash equilibria and quantal response equilibria of static games often are not expressable as closed-form formulae of the game parameters. Therefore, finding optimal parameters requires solving a bi-level optimization problem, which involves computing the directional derivatives of the equilibria with respect to game parameters by differentiating the nonlinear program that characterizes the game equilibria.
Several papers have studied differentiation through the nonlinear program. Gould et al. \cite{gould2016differentiating} describe the general techniques of differentiating possibly nonlinear optimization problems, but these optimization problems do not allow inequality constraints. There are also implicit-differentiation-based methods that express the equilibrium as an implicit function of game parameters using KKT matrices and then use matrix calculus \cite{magnus1988matrix} to derive the gradients \cite{amos2017optnet, amos2019differentiable, ralph1995directionalder}. More recently, these differentiable optimization methods appear in the context of game theory as well, both for differentiating through Nash equilibria \cite{liu2023unkonwnobj, peters2022learning}, and quantal response equilibria \cite{yu2023qrematrix}.

%% file: sections/3_preliminaries.tex
\section{Preliminaries}\label{sec:preliminaries}

\subsection{Static Games}
A \textit{static game} \( G = (N, S, c) \) is defined by three key elements. \( N = \{1, \ldots, n\} \) is a set of \(n\) players.
For each player \( i \in N \), there is a finite set of \textit{pure strategies}, or actions, \( S^i \) available. The set of all strategy profiles \( S = \prod_{i \in N} S^i \) is the Cartesian product of pure strategies for all players.
Each player \( i \) has a \textit{cost function} \( c^i: S \to \R \), which assigns a real number as a cost to each strategy profile. The tuple of these functions for all players is represented as \( c = (c^i)_{i \in N} \).

A \textit{mixed strategy} for player \( i \) is a probability distribution over their set of pure strategies \( S^i \), denoted as \( x^i \in \Delta_{|S^i|-1} \), where \( x^i(s) \) is the probability of choosing strategy \( s\in S^i \). Assuming elements in \(S^i\) are indexed from \(1\) to \(|S^i|\), each \( x^i \) can be represented as a non-negative vector in \( \mathbb{R}^{|S^i|} \), where \(\mathbbm{1}^\top x^i  = 1\). A \textit{mixed strategy profile} \( \mathbf{x} = (x^1, \ldots, x^n) \) is the tuple of mixed strategies of all \( n \) players.
For convenience, we introduce the notation \(s^{-i} \coloneqq (s^1, \dots, s^{i-1}, s^{i+1}, \dots s^{n})\) as the indexed set of pure strategies of all players except player \(i\). Similarly, \(x^{-i} \coloneqq (x^1, \dots,x^{i-1}, x^{i+1}, \dots, x^n)\) refers to the strategy profiles of all players other than player \(i\), which allows us to represent the mixed strategy profile for player \(i\) in the context of others as \( \mathbf{x}=(x^i, x^{-i}) \). 
Given a mixed strategy \(\mathbf{x}\) and the \(i\)'th player's cost function \(c^i\), the \textit{expected cost} of player \(i\) can be written as 
\begin{equation}
\begin{aligned}
     J(\mathbf{x},c^i) &\coloneqq \mathbb{E}_{s^1 \sim x^1, \ldots, s^n \sim x^n}\left[c^i(s^1,\ldots, s^n)\right]\\
     &=\sum_{s^1 \in S^1, \ldots, s^n \in S^n} \left(\prod_{i\in N}{x^i(s^i)}\right) c^i(s^1, \dots, s^n)
\end{aligned}
\label{eqn:J_expected_cost}
\end{equation}

This function is important as the objective functions of the optimization for both the forward game and the relationship design problems are in this form.

\subsection{Nash Equilibrium and Quantal Response Equilibrium}
The concept of \textit{Nash equilibrium (NE)} refers to a set of strategies where no player can benefit by unilaterally changing their strategy, given the strategies of the other players. Formally, in a static game \( G_c=(N,S,c) \), a mixed strategy profile \(\mathbf{x}=(x^i, x^{-i})\) is a Nash equilibrium if each player \(i\in N\) chooses the optimal strategy \(x^{i}\) given the strategies \(x^{-i}\) of the other players, i.e., \(\mathbf{x}=\mathrm{NE}(G_c)\) if and only if for all \( i\in N\),
\begin{equation}
\centering
\begin{aligned}
    \min_{x^i} \quad & J((x^i, x^{-i}), c^i)\\
    \textrm{s.t.} \quad & \mathbbm{1}^\top x^i =1\\
    \quad & x^i \ge 0
\end{aligned}
\label{eqn:nash_optimality_condition}
\end{equation}

The \textit{quantal response equilibrium (QRE)} concept builds on top of the optimality conditions for Nash in \cref{eqn:nash_optimality_condition} with an additional term of entropy for each player, i.e., \(\mathbf{x} = \QRE(G_c)\) if and only if for all $i \in N$,
\begin{equation}
\begin{aligned}
\min_{ x^i} \quad & J(\mathbf{x},c^i) - \lambda H(x^i)\\ 
\textrm{s.t.} \quad & \mathbbm{1}^\top {x^i} =1.
\end{aligned}
\label{eqn:entropy_nash_optimality_condition}
\end{equation}
Notice the non-negative constraints in \cref{eqn:nash_optimality_condition} are redundant since the logarithm function in the entropy term implies that \(x^i \ge 0\).

After rearranging we get for all $i \in N$,
\begin{equation}
    \begin{aligned}
    \min_{x^i} \quad & -\left(-\frac{1}{\lambda}J(x^{-i},c^i)\right)^\top x^i - H(x^i)\\ 
    \textrm{s.t.} \quad & \mathbbm{1}^\top {x^i} =1,
    \end{aligned}
\end{equation}
where we use the shorthand \(J(x^{-i}, c^i)_a\) as a vector whose \(a\)'th entry is given by,
\begin{equation}
    J(x^{-i},c^i)_a
        = \mathbb{E}_{s^{-i} \sim x^{-i}}\left[c^i(s^{-i}) | s^i = a\right].
\end{equation}
Intuitively, $J(x^{-i},c^i)$ denotes the vector of expected costs of player $i$ taking each possible action $a$ while all other players following the strategy profile $x^{-i}$.

Based on Theorem 4 in the thesis \cite{amos2019differentiable} by Amos, the solution for this optimization problem is in the form of a softmax function, where boundedly rational players choose strategies with probabilities proportional to their costs, i.e., $\mathbf{x} = \QRE(G_c)$ if and only if for all $i \in N$,
\begin{equation}
\begin{aligned}
    x^i &=\softmax\left(-\frac{1}{\lambda}J\left(x^{-i},c^i\right)\right),
\end{aligned}
\label{eqn:entropy_nash_optimality}
\end{equation}
where
\begin{equation}
\softmax(\mathbf{z})_i = \frac{e^{z_i}}{\sum_{j=1}^{n} e^{z_j}}.
\end{equation}

Inspired by \cite{yu2023qrematrix}, we approximately compute the quantal response equilibrium by solving the following nonlinear least-squares problem subject to the constraint that players' strategies need to be proper probability distributions.
\begin{equation}
\begin{aligned}
    & \underset{\mathbf{x}=(x^1,\ldots, x^n)}{\text{minimize}}
    & & \sum_{i\in N}\left\|x^i - \softmax\left(-\frac{1}{\lambda}J\left(x^{-i},c^i\right)\right)\right\|^2 \\
    & \text{subject to}
    & & \mathbbm{1}^\top {x^i} =1, \quad \forall i\in N
\end{aligned}
\label{eqn:least_squares}
\end{equation}

Essentially this optimization identifies a joint strategy for all players that maximally satisfies the quantal response equilibrium condition in \cref{eqn:entropy_nash_optimality}.

\subsection{Weighted Graphs}
In the study of strategic interactions among multiple players, \textit{graphs} can represent the complex existing relationships among players. Each graph is typically represented as an \textit{adjacency matrix} \( \phi \in \mathbb{R}^{n \times n} \), where \( n \) is the number of players and each matrix entry \( \phi_{ij}\in\{0,1\} \) indicates the presence of relationships from player \( i \) to player \( j \). In scenarios involving multiple types of relationships, a \textit{superposition} of such graphs effectively combines several adjacency matrices to create a comprehensive representation of all relationship dynamics. Assigning weights \(w_r\) on each adjacency matrix \(\phi_r\), we get the final combined network as \(\sum_{r=1}^k w_r\phi_r \in \mathbb{R}^{n\times n}\).

%% file: sections/4_problem.tex
\section{Relationship Weight Design Problems}\label{sec:problem}
We introduce a novel model called a \textit{relationship game} to capture the inter-agent relationships within a static game framework.

\noindent\begin{definition}[Relationship Game]
A \textit{relationship game} is an augmented static game $\mathcal{G}_c=(N,S,c,\Phi)$ that, in addition to the usual static game structure, contains an indexed set of relationships $\Phi = (\phi_1, \phi_2, \ldots, \phi_k)$ where each relationship $\phi_i \in \mathbb{R}^{n \times n}$ is an adjacency matrix representing a directed graph. A relationship game $\mathcal{G}_c$ together with a weight assignment $\mathbf{w} \in \mathbb{R}$ induces a static game $G_{\Tilde{c}} =(N, S, \Tilde{c})$, where the cost functions $\tilde c=(\tilde c^i)_{i\in N}$ are altered by \textit{modification functions} \(M(c, \Phi, w) = (M(c, \Phi, w)^i)_{i\in N}\) as follows,
\begin{equation}
\begin{aligned}
    & \tilde c^i(s) = M(c, \Phi, w)^i = \sum_{j\in N} \bigg(\sum_{r=1}^{k} w_r \phi_r \bigg)_{ij} c^j(s) \quad \forall s \in S.
\end{aligned}
\end{equation} 
\end{definition}
Intuitively, $\Tilde{c}^i$ is the weighted superposition of all neighbors of player $i$ in all possible relationship graphs. For convenience of reference, we write \(\tilde c = M(c,\Phi, \mathbf{w})\).



%

We specify the socially desirable behavior through a \textit{social cost function} $V: S\to \R$ that assigns real numbers to strategy profiles. Given a relationship game and a social cost function, the design problem aims to find a relationship weight vector $\mathbf{w}$ across relationship networks in $\Phi$ such that the quantal response equilibrium $\mathbf{x}^*$ of the induced game $G_{\tilde{c}}$ minimizes expected social cost $J(\mathbf{x}^*, V)$. This goal may not be well-defined if the game $G_{\tilde{c}}$ has multiple quantal response equilibria, in which case there can be more than one possible value for $J(\mathbf{x}^*, V)$. Thus, to ensure the problem is well-defined, we require a method to specify the equilibrium $x^*$ of $G_{\tilde{c}}$ on which to carry out the minimization of $J(\mathbf{x}^*, V)$. We propose two methods to do this. The first is to minimize the maximal attainable cost across all quantal response equilibria of $G_{\tilde{c}}$, and the second is to minimize the minimal attainable cost. Optimization problems \eqref{eqn:problem_minmax} and \eqref{eqn:problem_minmin} are the respective problems resulting from these two methods.

\begin{figure}[h]
    \centering
    \begin{minipage}{0.5\textwidth}
        \centering
        \includegraphics[width=\textwidth]{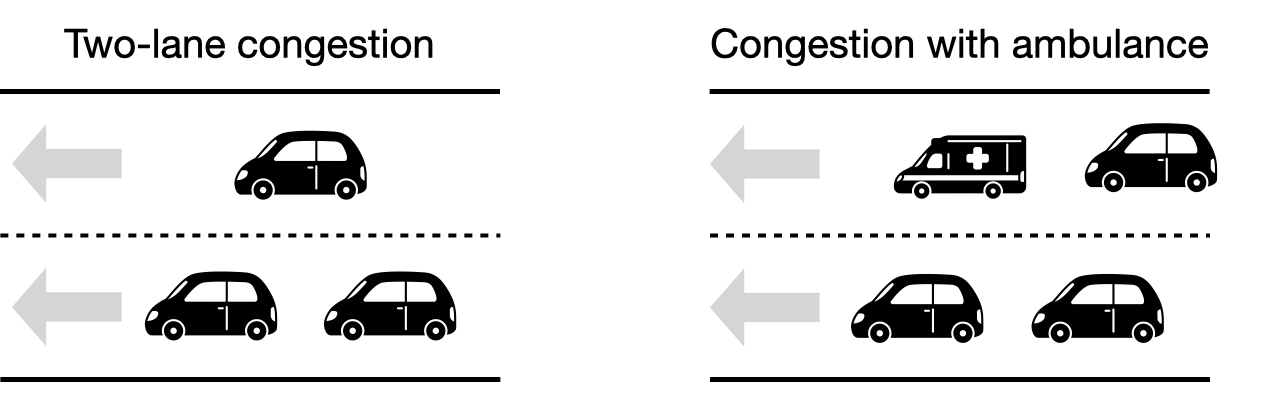}
        \caption{Descriptive illustrations for the two congestion game scenarios.}
        \label{fig:examples}
    \end{minipage}
    \hfill
    \begin{minipage}{0.45\textwidth}
        \centering
        \begin{tikzpicture}
            \begin{axis}[
                width=6cm, height=3.2cm,
                xlabel={\(w_3\)},
                ylabel={\(J(\mathbf{x}, V)\)},
                xmin=0, xmax=3,
                ymin=24, ymax=40,
                grid=major,
                grid style=dashed,
                legend style={
                    at={(0.5,1.4)},
                    anchor=north, 
                    legend columns=-1,
                    /tikz/every even column/.append style={column sep=2mm},
                    mark size=5pt
                }
            ]
        
            \addplot [only marks, mark options={color=black, scale=0.3}, forget plot] table [x=w_axis, y=data, col sep=comma]{eqs_data/eqs.dat};
            
            \addplot [only marks, mark options={color=blue, scale=0.3}] table [x=w_axis, y=data, col sep=comma]{eqs_data/eqs_min.dat};\label{plot:eqs_plot_min}
            \addlegendentry{Min}

            \addplot [only marks, mark options={color=orange, scale=0.3}] table [x=w_axis, y=data, col sep=comma]{eqs_data/eqs_max.dat};\label{plot:eqs_plot_max}
            \addlegendentry{Max}
            
            \addplot [only marks, mark options={color=violet, scale=0.3}] table [x=w_axis, y=data, col sep=comma]{eqs_data/eqs_equals.dat};\label{plot:eqs_plot_equals}
            \addlegendentry{Max=Min}
            
            \end{axis}
        \end{tikzpicture}
        \vspace{-0.6cm}
        \caption{Multiplicity of equilibria in \textit{congestion with ambulance} with \(\lambda=1.0\) and \(\mathbf{w}=[0.5, 0.5, w_3, 0.5]\), where \(w_3\) is sampled in the range \(0 \leq w_3 \leq 3\) with increments of 0.01. At each \(w_3\) value, we test \(300\) different seeds with a tolerance threshold of \(0.0001\) to identify unique values.}
        \label{fig:eqs_plot}
    \end{minipage}
\end{figure}

For both problems, the primary objective is to minimize a social cost function associated with a game equilibrium. However, as illustrated in \Cref{fig:eqs_plot}, multiple equilibria may emerge as the weight parameter $\mathbf{w}$ varies. The inner maximization/minimization thus selects an equilibrium,  parameterized by \(\sigma\), that has the highest/lowest social cost.
We define $\sigma$ as a real number parameter between $[0,1]$ and $\QRE(G_{\tilde{c}}, \sigma)$ is a surjective function that maps the entire range $[0,1]$ onto the set of all quantal response equilibria of $G_{\tilde{c}}$.

\begin{problem}[Relationship Weight Vector Design---Min-Max Form] 
\label{problem:weight_design_minmax}
Given a relationship game $\mathcal{G}_c=(N,S,c,\Phi)$ and a social cost function $V$,
\begin{equation}
\begin{aligned}
\min_{\mathbf{w}}\max_{\sigma} \quad & J(\mathbf{x}, V)\\
\textrm{s.t.} \quad & \Tilde{c} = M(c, \Phi, \mathbf{w})\\
  & \mathbf{x} = \QRE(G_{\tilde{c}}, \sigma) \\
\end{aligned}
\label{eqn:problem_minmax}
\end{equation}
\end{problem}

\begin{problem}[Relationship Weight Vector Design---Min-Min Form] 
\label{problem:weight_design_minmin}
Given a relationship game $\mathcal{G}_c=(N,S,c,\Phi)$ and a social cost function $V$,
\begin{equation}
\begin{aligned}
\min_{\mathbf{w}, \sigma} \quad & J(\mathbf{x}, V)\\
\textrm{s.t.} \quad & \Tilde{c} = M(c, \Phi, \mathbf{w})\\
  & \mathbf{x} = \QRE(G_{\tilde{c}}, \sigma) \\
\end{aligned}
\label{eqn:problem_minmin}
\end{equation}
\end{problem}

Here, \Cref{problem:weight_design_minmax} minimizes the maximal cost across all equilibria (among the set of equilibria marked in orange in \Cref{fig:eqs_plot}). Conversely, \Cref{problem:weight_design_minmin} minimizes the minimal cost across all equilibria which are marked in blue in the same figure. These two goals equate to minimizing the price of anarchy and the price of stability, respectively.

We explore the problems' relevance to two distinct scenarios of socially desirable behavior in the context of congestion games:

\begin{enumerate}
    \item \textbf{Two-lane Congestion}: This scenario involves a congestion game on a two-lane road. Each lane's delay is proportional to its vehicle count. The social cost function, representing total congestion time, is the aggregate of individual delay times. The objective is to minimize overall congestion time.
    \item \textbf{Congestion with Ambulance}: Similar to the two-lane case, but with an ambulance granted higher priority. The social cost function, a weighted sum of delays, prioritizes the ambulance. The goal is to facilitate ambulance passage, even if it increases delay for others.
\end{enumerate}

%% file: sections/5_method.tex
\section{Gradient-based Mechanism Design}
We outline how to solve the mechanism design problems \ref{problem:weight_design_minmax} and \ref{problem:weight_design_minmin} with a gradient-based optimization approach, which first requires a game solver to compute the quantal response equilibrium.
We implement the game solver by specifying the least-squares problem \eqref{eqn:least_squares} in Julia \cite{bezanson2017julia} using the JuMP \cite{dunning2017jump} interface and the COIN-OR IPOPT \cite{wachter2006implementation} optimizer.
Furthermore, we reroll \textit{seeds} to explore the parameter space of \(\sigma\).
Running with a different seed randomly samples a different value of parameter \(\sigma\), which sets a different random initial joint strategy vector \(\mathbf{x}\), possibly leading to a different quantal response equilibrium.
Since both the cost function \( \tilde c \) in the game and a parameter \(\sigma\) selected by a seed determines a game solution \(\mathbf{x}\), we succinctly represent this process as \(\mathbf{x} = \QRE(G_{\tilde c}, \sigma)\).

We now describe two tailored projected gradient descent algorithms, Min-Max and Min-Min, to solve the mechanism design problems \ref{problem:weight_design_minmax} and \ref{problem:weight_design_minmin}, respectively.
Both algorithms share a common set of inputs: a relationship game \(\mathcal{G}_c=(N,S,c,\Phi)\), a social cost function \(V\), a step size \(\alpha\), a convergence threshold \(\beta\), and a reroll number \(L\). 
Both aim to determine an optimal relationship weight vector \(\mathbf{w}\), though with different objectives: 
\begin{itemize}
    \item Min-Max identifies a \(\mathbf{w}\) that provides the lowest possible upper bound on \(J(\mathbf{x}, V)\);
    \item Min-Min seeks the lowest possible value of \(J(\mathbf{x}, V)\) associated with \(\mathbf{w}\).
\end{itemize}

{\noindent\bf Gradient Descent.}
Recall from \Cref{problem:weight_design_minmax} we define the objective function, \(J(\mathbf{x}, V)\), in terms of the social cost for the modified game solution, where \(\mathbf{x} = \QRE(G_{\tilde{c}}, \sigma)\) and \(\tilde{c} = M(c, \Phi, \mathbf{w})\). Since \(c, \Phi, V\) are given parameters in this problem, the objective function is parameterized only by a relationship weight vector \(\mathbf{w}\) and a seed-selected value  \(\sigma\). Hence we simplify the notation to express the social cost as a function of \(\mathbf{w}\) and \(\sigma\), denoted by \(J(\mathbf{w}, \sigma)\). Given the step size \(\alpha\), we update the weight vector as
\begin{equation}
    \mathbf{w} - \alpha\nabla_{\mathbf{w}}J(\mathbf{w},\sigma).
\end{equation}
We refer the readers to Appendix \ref{sec:gradient} for the derivation of \(\nabla_{\mathbf{w}}J(\mathbf{w},\sigma)\), the gradient of \(J(\mathbf{w},\sigma)\) with respect to \(\mathbf{w}\).

{\noindent\bf \(L_2\) Projection.}
After the gradient step, we project the output \( \|\mathbf{w}\| \) onto the unit sphere as follows.
\begin{equation}
    \text{Proj}_{L_2}(\mathbf{w}) = \frac{\mathbf{w}}{\|\mathbf{w}\|}.
\end{equation}
This projection enforces the constraint in \cref{eqn:entropy_nash_optimality_condition}. Furthermore, we employ this projection for two reasons. Firstly, it allows us to constrain the range of \( \mathbf{w} \) and prevent them from diverging to infinity, which allows the gradient descent to converge faster. Secondly, constraining \( \|\mathbf{w}\| \) prevents \(J(\mathbf{w},\sigma)\) from taking arbitrarily small or large values, which can cause the bounded rationality term \(\lambda H(x^i)\) to either dominate the game objective or be negligible, defeating the purpose of modeling the agent behavior using a quantal response equilibrium.

{\noindent\bf Convergence condition.} 
After computing the gradient in each iteration, we use the condition below to check for convergence:
\begin{equation}
    \left\|\nabla_\mathbf{w} J(\mathbf{w},\sigma) - \frac{\nabla_\mathbf{w} J(\mathbf{w},\sigma)^\top \mathbf{w}}{\|\mathbf{w}\|^2} \mathbf{w}\right\| < \beta.
\end{equation}
Intuitively, this condition ensures that the directional derivatives that lie tangentially to the unit sphere are smaller than a threshold parameter $\beta$, implying that the weights \(\mathbf{w}\) are near a critical point of \(J(\mathbf{w},\sigma)\) on the unit \(L_2\) sphere.

{\noindent\bf Min-Max.}
Recall that we may have multiple equilibria in a game, and we can switch to different branches with different seeds in the game solver. Thus to only search within the top branch, we reroll seeds to find a seed that provides maximal social cost. Then the algorithm commits to this seed and performs the gradient computation.

{\noindent\bf Min-Min.}
The idea in this algorithm is to keep social cost non-increasing, and retry with a different seed otherwise.
We take a projected gradient descent step and keep the new weight vector as a candidate \(\overline{\mathbf{w}}\), then we only update \(\mathbf{w}\) when its social cost is smaller or equal to the previous social cost, else we increment the seed number and continue to the next iteration.

\begin{minipage}{.46\textwidth}
    \begin{algorithm}[H]
    \caption{Min-Max Projected Gradient Descent}\label{algo:minmax}
    \begin{algorithmic}
    \State Initialize \(k=1, \mathbf{w}\sim Uniform(|\Phi|)\)
    \While{\(k<k_{max}\)}
        \State \(\hat{J}=-\infty\)
        \For{\(\sigma_{\text{test}} \in [0,1]\)}
            \If{\(J(\mathbf{w}, \sigma_{\text{test}}) > \hat{J}\)}
                \State \(\sigma \gets \sigma_{\text{test}}\)
                \State \(\hat{J} \gets J(\mathbf{w}, \sigma_{\text{test}})\)
            \EndIf
        \EndFor
        \State Compute gradient \(\nabla_\mathbf{w} J(\mathbf{w},\sigma)\)
        \State \(\mathbf{w} \gets \text{Proj}_{L_2}(\mathbf{w} - \alpha \nabla_\mathbf{w} J(\mathbf{w},\sigma))\)
        \If{\(\left\|\nabla_\mathbf{w} J(\mathbf{w},\sigma) - \frac{\nabla_\mathbf{w} J(\mathbf{w},\sigma)^\top \mathbf{w}}{\|\mathbf{w}\|^2} \mathbf{w}\right\| < \beta\)}
            \State terminate with \(\mathbf{w}\).
        \EndIf
        \State \(k\gets k+1\)
    \EndWhile
    \end{algorithmic}
    \end{algorithm}
\end{minipage}%
\hfill
\begin{minipage}{.46\textwidth}
    \begin{algorithm}[H]
    \caption{Min-Min Projected Gradient Descent}\label{algo:downstairs}
    \begin{algorithmic}
    \State Initialize \(k=1, \mathbf{w}\sim Uniform(|\Phi|)\)
    \State Set \(\sigma=0, J_{\text{prev}} = J(\mathbf{w},\sigma)\)
    \While{\(k<k_{max}\)}
        \State Compute gradient \(\nabla_\mathbf{w} J(\mathbf{w},\sigma)\)
        \If{\(\left\|\nabla_\mathbf{w} J(\mathbf{w},\sigma) - \frac{\nabla_\mathbf{w} J(\mathbf{w},\sigma)^\top \mathbf{w}}{\|\mathbf{w}\|^2} \mathbf{w}\right\| < \beta\)}
            \State terminate with \(\mathbf{w}\).
        \EndIf
        \State \(\overline{\mathbf{w}} = \text{Proj}_{L_2}(\mathbf{w} - \alpha \nabla_\mathbf{w} J(\mathbf{w},\sigma))\)
        \If{\(J(\overline{\mathbf{w}}, \sigma) > J_{\text{prev}}\)}
            \State \(\sigma \gets \sigma + \frac{1}{L}\), continue.
        \Else
            \State \(\mathbf{w}\gets \overline{\mathbf{w}}\)
            \State \(J_{\text{prev}} = J(\mathbf{w},\sigma)\)
        \EndIf
        \State \(k\gets k+1\)
    \EndWhile
    \end{algorithmic}
    \end{algorithm}
\end{minipage}
\vspace{2em}

We implement the two algorithms in Julia, calling the game solver described in the beginning of this section. The code of both the game solver and the two projected gradient descent algorithms are publicly available at \url{https://github.com/vivianchen98/relationship_game}.

%% file: sections/6_numerical.tex
\section{Numerical Examples}\label{sec:numerical}
We test the proposed algorithms in the two congestion game examples described at the end of \Cref{sec:problem}.

In \textbf{two-lane congestion}, we consider three players traveling on a road with two lanes, denoted as \(a\) and \(b\). The \textit{load}, or the number of players choosing this lane, are \(l_a\) and \(l_b\). Each player chooses between these two lanes. The delay a player experiences in a lane is determined by that lane's load: \(l_a\) in lane \(a\) and \(2 \times l_b\) in lane \(b\).
The cost for each player is based on the delay they experience, influenced by their own lane choice as well as the choices of other players. 
The social cost in this example is the sum of all players' costs: \(V(S) = \sum_{i=1}^3 c^i(S)\).
The indexed set of relationships includes an identity matrix to account for each player's individual cost and a directed matrix capturing the relation from each player to the other two players.
\begin{equation}
    \Phi = \left\{
    \underbracket[0.6pt]{\begin{bmatrix}
        1 & 0 & 0 & \\
        0 & 1 & 0 & \\
        0 & 0 & 1 & 
    \end{bmatrix}}_\text{selfish}, \;
    \underbracket[0.6pt]{\begin{bmatrix}
        & 0 & 1 & 1 \\
        & 1 & 0 & 1 \\
        & 1 & 1 & 0 
    \end{bmatrix}}_\text{to others}
    \right\}
\end{equation}

In \textbf{congestion with ambulance}, we introduce an additional ambulance (\textbf{A}) into the two-lane congestion example of regular cars (\textbf{R}), with the same load functions.
Here, the social cost is a weighted sum of the costs incurred by all players, with a higher weight on the ambulance to prioritize its movement:
\begin{equation}
    V(S) = 8\cdot c^{\textbf{A}}(S) + \sum_{i\in \textbf{R}} c^i(S)
\end{equation}
The relationship basis includes an identity matrix for each player's individual cost, along with matrices representing interactions among regular cars, from regular cars to the ambulance, and from the ambulance to regular cars.
\begin{equation}
    \Phi = \left\{
        \underbracket[0.6pt]{\begin{bmatrix}
            1 & 0 & 0 & 0 \\
            0 & 1 & 0 & 0 \\
            0 & 0 & 1 & 0 \\
            0 & 0 & 0 & 1
        \end{bmatrix}}_\text{selfish},\;
        \underbracket[0.6pt]{\begin{bmatrix}
            0 & 0 & 0 & 0 \\
            0 & 0 & 1 & 1 \\
            0 & 1 & 0 & 1 \\
            0 & 1 & 1 & 0
        \end{bmatrix}}_\text{among \textbf{R}},\;
        \underbracket[0.6pt]{\begin{bmatrix}
            0 & 0 & 0 & 0 \\
            1 & 0 & 0 & 0 \\
            1 & 0 & 0 & 0 \\
            1 & 0 & 0 & 0
        \end{bmatrix}}_\text{\textbf{R} to \textbf{A}},\;
        \underbracket[0.6pt]{\begin{bmatrix}
            0 & 1 & 1 & 1 \\
            0 & 0 & 0 & 0 \\
            0 & 0 & 0 & 0 \\
            0 & 0 & 0 & 0
        \end{bmatrix}}_\text{\textbf{A} to \textbf{R}}
    \right\}
\end{equation}

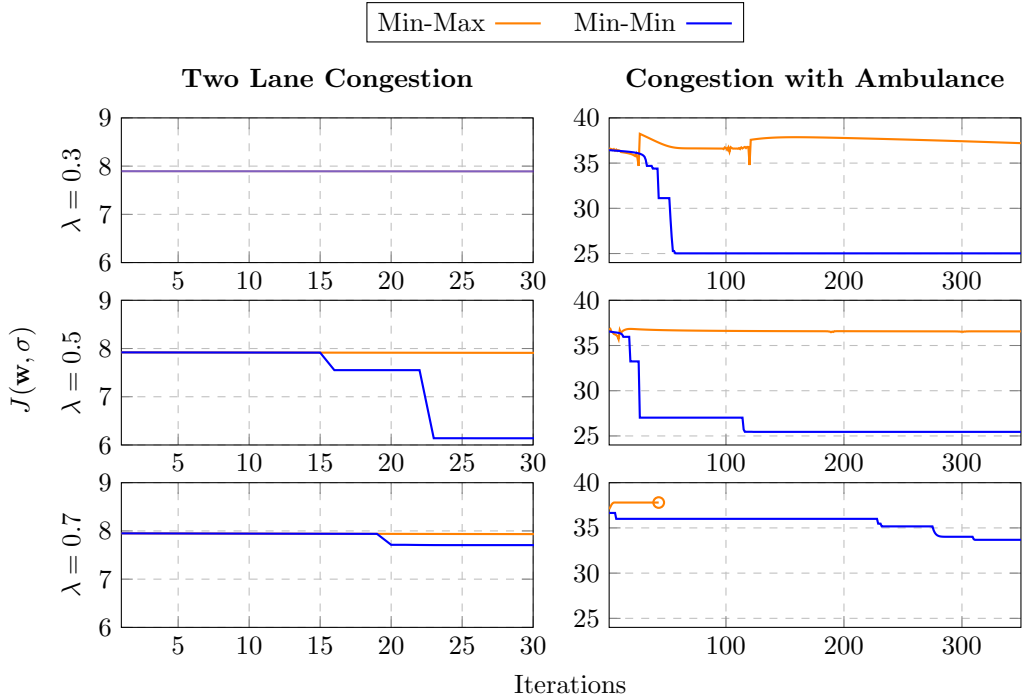
\begin{figure}[ht]
    \centering
    \begin{tikzpicture}
        \begin{groupplot}[
            group style={
                group size=2 by 3, 
                vertical sep=0.5cm, 
                horizontal sep=1cm, 
            },
            width=7cm, height=3.5cm, 
            grid=major,
            grid style=dashed,
            xmin=1, xmax=1000,
        ]
        
        \nextgroupplot[title=\textbf{Two Lane Congestion}, xmin=1, xmax=30, ymin=6, ymax=9, ylabel={$\lambda=0.3$}]
        \addplot [mark=none, orange, thick, opacity=0.5] table [x=iter, y=minmax, col sep=comma]{convergence_data/congestion_0.3.dat};\label{plot:congestion_0.3_minmax}
        \addplot [mark=none, blue, thick, opacity=0.5] table [x=iter, y=downstairs, col sep=comma]{convergence_data/congestion_0.3.dat};\label{plot:congestion_0.3_downstairs}

        \nextgroupplot[title=\textbf{Congestion with Ambulance}, xmin=1, xmax=350, ymin=24, ymax=40]
        \addplot [mark=none, orange, thick] table [x=iter, y=minmax, col sep=comma]{convergence_data/bee_queen_0.3.dat};\label{plot:bee_queen_0.3_minmax}
        \addplot [mark=none, blue, thick] table [x=iter, y=downstairs, col sep=comma]{convergence_data/bee_queen_0.3.dat};\label{plot:bee_queen_0.3_downstairs}

        \nextgroupplot[xmin=1, xmax=30, ymin=6, ymax=9, ylabel={$\lambda=0.5$}]
        \addplot [mark=none, orange, thick] table [x=iter, y=minmax, col sep=comma]{convergence_data/congestion_0.5.dat};\label{plot:congestion_0.5_minmax}
        \addplot [mark=none, blue, thick] table [x=iter, y=downstairs, col sep=comma]{convergence_data/congestion_0.5.dat};\label{plot:congestion_0.5_downstairs}

        \nextgroupplot[xmin=1, xmax=350, ymin=24, ymax=40]
        \addplot [mark=none, orange, thick] table [x=iter, y=minmax, col sep=comma]{convergence_data/bee_queen_0.5.dat};\label{plot:bee_queen_0.5_minmax}
        \addplot [mark=none, blue, thick] table [x=iter, y=downstairs, col sep=comma]{convergence_data/bee_queen_0.5.dat};\label{plot:bee_queen_0.5_downstairs}

        \nextgroupplot[xmin=1, xmax=30, ymin=6, ymax=9, ylabel={\textbf{$\lambda=0.7$}}]
        \addplot [mark=none, orange, thick] table [x=iter, y=minmax, col sep=comma]{convergence_data/congestion_0.7.dat};\label{plot:congestion_0.7_minmax}
        \addplot [mark=none, blue, thick] table [x=iter, y=downstairs, col sep=comma]{convergence_data/congestion_0.7.dat};\label{plot:congestion_0.7_downstairs}

        \nextgroupplot[xmin=1, xmax=350, ymin=24, ymax=40]
        \addplot [mark=none, orange, thick] table [x=iter, y=minmax, col sep=comma]{convergence_data/bee_queen_0.7.dat};\label{plot:bee_queen_0.7_minmax}
        \addplot [mark=none, blue, thick] table [x=iter, y=downstairs, col sep=comma]{convergence_data/bee_queen_0.7.dat};\label{plot:bee_queen_0.7_downstairs}
        \addplot[only marks, mark=o, orange, thick]coordinates {(43,37.804660943020366)};
        
        \end{groupplot}

        \node[rotate=90, anchor=south] at ([xshift=-1cm, yshift=1cm]group c1r2.south west) {\(J(\mathbf{w}, \sigma)\)}; 
        \node[anchor=north] at ([xshift=3.2cm, yshift=-0.5cm]group c1r3.south) { Iterations}; 
        
        \matrix[
            matrix of nodes,
            anchor=south,
            draw,
            inner sep=0.2em,
            column sep=0.8em,
        ] (legend) at ([xshift=3cm, yshift=1cm]group c1r1.north) { 
            Min-Max \ref{plot:congestion_0.5_minmax} & Min-Min \ref{plot:congestion_0.5_downstairs} \\        
        };
    \end{tikzpicture}
\caption{The convergence of the iterates in \Cref{algo:minmax} (Min-Max) and \Cref{algo:downstairs} (Min-Min) on both congestion game examples. Termination points are highlighted with circles. The plots show the social cost at the current equilibrium \(J(\mathbf{w}, \sigma) = J(\mathbf{x}, V)\) with respect to the iteration number.}
\label{fig:convergences}
\end{figure}

We demonstrate the effects of applying Min-Max and Min-Min projected gradient descent on the two examples. Throughout, we let \(\alpha=0.1\),  \(\beta=0.0001\), and the maximum number of iterations be \(2000\).

\begin{figure}[ht!]
    \centering
    \includegraphics[width = \textwidth]{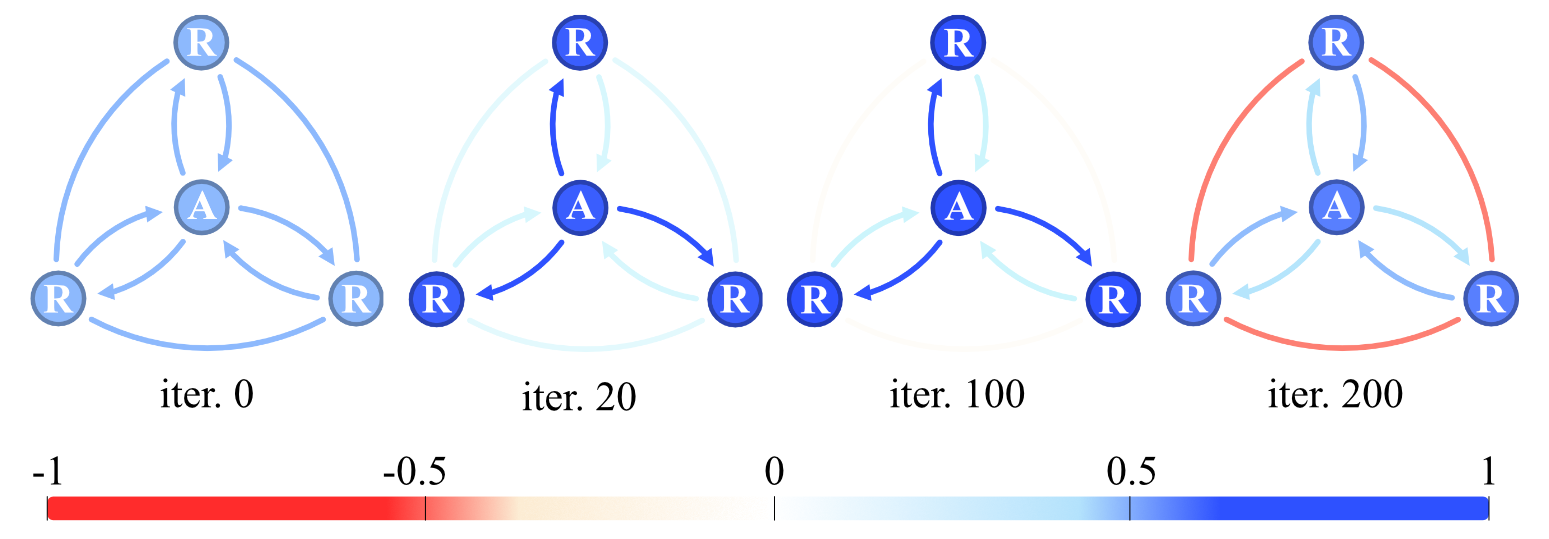}
    \caption{Superposed relationship graphs for selected iterations of the Min-Min optimization for the congestion with ambulance example, where $\lambda = 0.5$. Node colors represent the identity matrix for selfish costs, while undirected lines indicate bidirectional relationships between regular cars (\textbf{R}s).}
    \label{fig:network_graph}
\end{figure}

\Cref{fig:convergences} shows the convergence of both algorithms for the two examples, with the entropy weight vector \(\lambda\) varied across \{0.3, 0.5, 0.7\}. For a \(\lambda\) value of 0.3 in the two-lane congestion scenario, the Min-Max algorithm coincides with the Min-Min algorithm, indicating the presence of a single equilibrium under this condition. 
As expected, the Min-Min algorithm consistently finds the lowest social cost, representing the best possible interaction outcome, while the Min-Max algorithm identifies the ``upper bound" of social cost among all potential equilibria. This does not necessarily yield the most optimal interaction outcome, but it ensures the final game solution derived from the weight vector is at least as good as that particular social cost value. 
Due to random seed rerolls, Min-Max can occasionally fail to find the equilibrium with the highest social cost. This is reflected as the spikes in orange lines for \(\lambda=0.3, 0.5\) in the congestion with ambulance example. Setting the number of rerolls to be high alleviates this problem at the cost of runtime.
As we increase the entropy weight, the Min-Min algorithm converges towards higher social cost values because higher entropy introduces more randomness to the game solution and worsens the social cost.
Min-Max terminates early in congestion with ambulance with \(\lambda=0.7\) as it quickly steps into an equilibrium which has a terminating condition value below the set threshold \(0.0001\), while the other trials do not meet the condition early on.

In the congestion with ambulance scenario, \Cref{fig:network_graph} demonstrates the evolution of the weighted inter-agent relationships through network graphs. Initially, the algorithm assigns equal weights to all relationships. Over iterations, the regular cars adapt by increasingly prioritizing the ambulance (indicated by a gradually darkening blue line from \textbf{R} to \textbf{A}), while competing among themselves (shown as red bidirectional lines between \textbf{R}s). This visualization highlights how the evolved weight vector prompts regular cars to exhibit altruistic behavior towards the ambulance, mirroring the real-world social norm of yielding to emergency vehicles. The graph at iteration \(200\) reveals a counter-intuitive result: the proposed algorithm determines it is optimal for regular cars to actively impede each other by negatively incorporating each other's costs, in order to prioritize the passage of the ambulance.

%% file: sections/7_conclusion.tex
\section{Conclusion}\label{sec:conclusion}
In this paper, we present a novel approach to model socially-aware decision-making in autonomous agents within a static game framework. We first define the concept of a relationship game, where an indexed set of player relationships augments a static game. We formulate two problems to optimize the weight vector corresponding to this set of relationships such that the modified game solution minimizes an objective function: the minimal social cost for Min-Min and the maximal social cost for Min-Max. We then propose two projected gradient descent algorithms tailored to solve each of these two problems and demonstrate their effectiveness numerically in two congestion game scenarios.

A limitation of our approach is the use of stochastic algorithms, which inherently require a balance between the number of seeds and runtime efficiency. Increasing the number of seeds can enhance the robustness and accuracy of the solution, but this comes at the expense of increased computational time. This trade-off can be particularly challenging in scenarios where rapid decision-making is critical, or when computational resources are limited. Future work could focus on developing more efficient stochastic methods that require fewer seeds to achieve a similar level of accuracy.

%% file: sections/appendix_inverse.tex
\section{Gradient Derivation via Implicit Differentiation}\label{sec:gradient}
Let \(\mathbf{z}\) be the flattened version of \(\mathbf{x}\), i.e., 
\[\mathbf{z}=f(\mathbf{x})=\begin{bmatrix}{x^1}^\top {x^2}^\top \ldots {x^n}^\top\end{bmatrix}^\top.\]


\(f\) induces a natural bijection between \(\mathbf{x}\) and \(\mathbf{z}\). Thus with a slight abuse of notation, we define
\begin{equation}
    J(\mathbf{z},V) \coloneqq J\left(f^{-1}(\mathbf{z}), V\right) = J(\mathbf{x}, V).
\end{equation}

The gradient \(\nabla_\mathbf{w} J(\mathbf{x}, V)\) we want can instead be computed by
\begin{equation}
    \begin{aligned}
        \nabla_\mathbf{w} J(\mathbf{z}, V) &= \nabla_\mathbf{z}J(\mathbf{z},V)\cdot \frac{\partial \mathbf{z}}{\partial \mathbf{w}}.
    \end{aligned}
\end{equation}

The first gradient of expected cost with respect to \(\mathbf{x}\) is straightforward:
\begin{equation}
    \begin{aligned}
        \nabla_\mathbf{z}J(\mathbf{z},V) 
        &= \nabla_{f(\mathbf{x})} J\left(f(\mathbf{x}), V\right) = f\left(\nabla_{\mathbf{x}} J(\mathbf{x}, V)\right)
        = \left[ \frac{\partial J}{\partial x^i_j} \right]_{\substack{i \in N, \\ \forall j \in S^i}} \text{where}\\
        \frac{\partial J}{\partial x^i_j} &= \sum_{\substack{s^i=j, \\ s^{-i}\in S^{-i}}} \left(V(\mathbf{s})\prod_{t\neq i}x^{t}(s^{t})\right).
    \end{aligned}
\end{equation}

The partial derivative \(\frac{\partial \mathbf{z}}{\partial \mathbf{w}} \) can be derived by applying the implicit function theorem on the KKT condition of Nash equilibrium. 

We define \(h^i(\mathbf{w},\mathbf{z})\) to be the vector of expected utilities of player \(i\) based on its choice of pure strategy. Similarly, \(g^{i,j}(\mathbf{w},\mathbf{z})\) is the matrix of expected utilities of player \(i\) with respect to the pure strategies of players \(i\) and \(j\). That is,
\begin{equation}
    \begin{aligned}
        h^i(\mathbf{w}, \mathbf{z}) 
        &= h^i\left(\mathbf{w}, f(\mathbf{x})\right) 
        = \left[\sum_{\substack{s^i=r, \\ s^{-i}\in S^{-i}}} \left(\tilde{c}^i(\mathbf{s}) \prod_{t\neq i}x^{t}(s^{t})\right)\right]_r, \\
        g^{i,j}(\mathbf{w}, \mathbf{z}) 
        &= g^{i,j}\left(\mathbf{w}, f(\mathbf{x})\right)
        = \left[\sum_{\substack{s^i=r, s^j = c, \\ s^{-i,j}\in S^{-i,j}}} \left(\tilde{c}^i(\mathbf{s})\prod_{t\neq i, j}x^{t}(s^{t})\right)\right]_{rc}.
    \end{aligned}
\end{equation}

Then we let
\[ F(\mathbf{w}, \mathbf{z}) = \mathbf{z} - s(\mathbf{w}, \mathbf{z}) = 0, \quad \text{where} \quad s(\mathbf{w}, \mathbf{z}) = \left[\softmax\left(-\frac{1}{\lambda}h^i(\mathbf{w}, \mathbf{z})\right)\right]_{i\in N},\]
and by implicit function theorem we have
\begin{equation}
    \frac{\partial \mathbf{z}}{\partial \mathbf{w}} = - \left(\frac{\partial F}{\partial \mathbf{z}}\right)^{-1} \frac{\partial F}{\partial \mathbf{w}}.
\label{eqn:implicit_function}
\end{equation}

With $|\mathbf{z}| = \underset{i\in N}{\sum}|S^i|$ and \( \mathbb{J}_{\softmax}\) as the Jacobian of the softmax function, we have
\begin{equation}
\begin{aligned}
    \frac{\partial F}{\partial \mathbf{z}} &= \frac{\partial \mathbf{z}}{\partial \mathbf{z}} - \frac{\partial s}{\partial \mathbf{z}} = I_{|\mathbf{z}|} + \frac{1}{\lambda} \mathbb{J}_{\softmax}\left(-\frac{1}{\lambda} h^i(\mathbf{w}, \mathbf{z})\right) g^{i,j}(\mathbf{w}, \mathbf{z}),\\
    \frac{\partial F}{\partial \mathbf{w}} &= - \frac{\partial s}{\partial \mathbf{w}} = \frac{1}{\lambda} \mathbb{J}_{\softmax}\left(-\frac{1}{\lambda} h^i(\mathbf{w}, \mathbf{z})\right) \frac{\partial h^i}{\partial \mathbf{w}},
\end{aligned}
\end{equation}
where \(\dfrac{\partial h^i}{\partial \mathbf{w}}\) is the Jacobian of \(h_i\) with respect to \(\mathbf{w}\). That is,
\begin{equation}
    \frac{\partial h^i}{\partial \mathbf{w}} = \left[\underset{\substack{s^i=r,\\ s^{-i}\in S^{-i}}}{\sum} \left(\sum_{j\in N} \phi^c_{ij} c^j(\mathbf{s})\right) \prod_{t\ne i}{x^t(s^t)}\right]_{rc}.
\end{equation}

In the implementation, Julia's 
\(\backslash\) operator, performing QR factorization, computes \cref{eqn:implicit_function} with better numerical stability.